\documentclass[aps,nofootinbib,longbibliography,twocolumn]{revtex4-1}

\usepackage{amsmath,amssymb,mathrsfs}
\usepackage{bbm} 
\usepackage[dvips]{graphicx}
\usepackage[dvips,bookmarks=false]{hyperref} 
\hypersetup{pdfstartview=FitH,pdfhighlight=/O,colorlinks=false}

\newcommand{\mbb}[1]{\mathbb{#1}}

\newcommand{\comment}[1]{}

\begin{document}
\title{\Large \bf Local spin foams}

\author{Elena Magliaro}  
\email{magliaro@gravity.psu.edu}
\author{Claudio Perini}     
\email{perini@gravity.psu.edu}
\affiliation{Institute for Gravitation and the Cosmos, Physics Department, Penn State, University Park, PA 16802-6300, U.S.A.}
\date{\small\today}

\begin{abstract}\noindent 
The central object of this paper is a holonomy formulation for spin foams. Within this new representation, we analyze three general requirements: locality, composition law, cylindrical consistency.  In particular, cylindrical consistency is shown to fix the arbitrary normalization of the vertex amplitude in the case of Euclidean signature. 
\end{abstract}

\maketitle
\emph{Dedicated to our families, overseas.}
\section{Introduction}
In this paper we consider the \emph{holonomy representation} for spin foams \cite{Bianchi:2010mw}. This representation allows to write spin foams in a Feynman path-integral form, where the configuration variables are $SU(2)$ group elements.

Spin foam models \cite{Reisenberger:2000fy,Perez:2003vx,Livine:2007vk,Engle:2007wy,Freidel:2007py,Ashtekar:2010ve} provide the transition amplitude from an `in' state to an `out' state of 3-geometry: they give a mathematical and physical meaning to the formal expression
\begin{equation}
W[g^{\text{\tiny $(3)$}}_{in},g^{\text{\tiny $(3)$}}_{out}]=\int_{g^{\text{\tiny $(3)$}}_{in}}^{g^{\text{\tiny $(3)$}}_{out}} D g^{\text{\tiny $(4)$}}\; \exp i S[g^{\text{\tiny $(4)$}}]
\end{equation}
for the Misner-Hawking transition amplitude of $3$-geometries in terms of a sum over $4$-geometries \cite{Misner:1957wq,Hawking:1979zw}. The recent convergence between covariant and canonical approaches to quantum gravity strengthened the idea that the spin foam theory constitutes a good alternative framework for the dynamics of Loop Quantum Gravity. In particular: i) in the 'new' spin foam models \cite{Livine:2007vk,Engle:2007wy,Freidel:2007py,Ding:2010ye} the correct weak imposition of linearized simplicity constraints provides a kinematical boundary Hilbert space isomorphic to the one of Loop Quantum Gravity. Hence we can interpret a spin foam amplitude as a transition amplitude between Loop Quantum Gravity kinematical states. This is also the philosophy behind the computation of n-point functions \cite{Rovelli:2005yj,Bianchi:2006uf,Alesci:2007tx,Bianchi:2009ri}. ii) the $SU(2)$ formulation of EPRL spin foam model \cite{Engle:2007wy,Bianchi:2010mw} respects a composition rule, studied in \cite{Bianchi:2010fj}, typical of canonical dynamics.  iii) here we make a further step: we introduce cylindrical consistency in spin foams. This is required for the interpretation of covariant amplitudes as transitions between wave-functions of a connection.  The last step goes in the direction of defining spin foam dynamics for the full LQG Hilbert space \cite{Kaminski:2009fm,Kaminski:2009cc}, and not for a truncation of it on a fixed (simplicial) graph. 

The paper is organized as follows. Section II is a review of the holonomy local formulation for spin foams. In section III we discuss the locality and composition properties (studied in \cite{Bianchi:2010fj}) in this formulation, and introduce the requirement of cylindrical consistency in spin foams. In particular we analyze invariance under face orientation reversal, face splitting, and face erasing. Similar invariance properties were considered in \cite{Baez:1997zt,Bojowald:2009im}, as an implementation of diffeomorphism invariance and in analogy with cylindrical consistency. Here our perspective is different: we require consistency to make a deeper contact with Loop Quantum Gravity. As a technical byproduct, we are able to extend the projection map on the solution of simplicity constraints to the full Hilbert space. Two subsections are dedicated to the specific cases of Ponzano-Regge and EPRL spin foams, both in the Euclidean signature. The Lorentzian signature is not studied in this paper and we leave it for a future investigation.

Section IV is a short discussion on the relationship between continuum limit and cylindrical consistency.
\section{Holonomy formulation}
In Loop Quantum Gravity \cite{Ashtekar:1986yd,Rovelli:1989za,Ashtekar:1991kc,Rovelli:2004tv,Thiemann:2007zz}, the kinematical Hilbert space is attached to a diffeomorphism-equivalence class of graphs embedded in a 3-dimensional space-like Cauchy hypersurface $\Sigma$. For a given equivalence class $\Gamma$, it is the Hilbert space $\mathcal{H}_\Gamma=L^2(SU(2)^L/SU(2)^N)$ where $L$ is the number of links of the graph and $N$ the number of nodes, so a state is a gauge invariant function of $SU(2)$ group elements $h_l\;$ ($l=1\ldots L$) that is invariant under $SU(2)$ gauge transformations at nodes, 
\begin{equation}
\Psi(h_l)=\Psi(g_{s(l)}\, h_l\, g^{-1}_{t(l)}).
\label{eq:gauge inv}
\end{equation}
Here $s(l)$ and $t(l)$ are respectively the nodes which are source/target of the link $l$, according to the orientation of the link. The full Hilbert space \cite{Ashtekar:1993wf} of loop gravity 
\begin{align}
\mathcal H=L^2(\overline{\mathcal A},d\mu_{AL})
\label{fullLQG}
\end{align}
is separable \cite{Fairbairn:2004qe} thanks to spatial diffeomorphism invariance\footnote{The group of diffeomorphisms must be extended to allow for isolated points in which maps are not differentiable, but still continuous.}, and decomposes into the orthogonal sum
\begin{align}
\mathcal H=\bigoplus_\Gamma \mathcal H_\Gamma.
\end{align}
By the Peter-Weyl theorem, an orthonormal basis of $\mathcal H_\Gamma$ is given by spin-network functions:
\begin{align}
\psi_{\Gamma j_l i_n}(g_l)=\otimes_n v_{i_n}\cdot\otimes_l \sqrt{2j_l+1}\,D^{j_l}(g_l)
\end{align}
labeled by spins $j_l$ (one per each link $l$) and intertwiners $i_n$ (one per each node $n$); $i_n$ is the label of an orthonormal basis $v_{i_n}$ in the space of intertwiners. The pattern of the contraction map ``$\cdot$'' is determined from the graph $\Gamma$. Finally $D^{j}$ is the Wigner $SU(2)$ representation matrix.

For a given $\mathcal H_\Gamma$, the configuration variables $h_l$ are interpreted as holonomies of the Ashtekar-Barbero connection \cite{Barbero:1994ap,Immirzi:1996di}
\begin{align}
A_a^i=\Gamma_a^i+\gamma K_a^i
\end{align}
along the link $l$ of the graph ($\Gamma_a^i$ is the spin-connection, $K_a^i$ the extrinsic curvature of the hypersurface $\Sigma$, and the real number $\gamma\neq 0$ is the Barbero-Immirzi parameter).

The partition function of a spin foam model takes the form of a sum\footnote{The sum over 2-complexes can be generated by an auxiliary Group Field Theory \cite{Oriti:2009wn}. It is usually not well-defined (divergent), and requires suitable gauge-fixing or regularization. For a fixed 2-complex, divergencies are associated to ``bubbles''. In topological (i.e. unphysical) theories, topological invariance implies that those ``bubbles'' can be removed up to a divergent overall factor that depends only on the cutoff. In quantum general relativity the situation is different: ``bubble'' divergencies are true radiative corrections which carry information about the infrared behavior of the theory \cite{Perini:2008pd,Freidel:2009hd,Bonzom:2010zh,Krajewski:2010yq}.} over partition functions for 2-complexes $\sigma$
\begin{equation}
Z=\sum_{\sigma} Z_\sigma.
\label{Zeta}
\end{equation}
A 2-complex is a collection of 2-dimensional \emph{faces}, 1-dimensional \emph{edges} and 0-dimensional \emph{vertices}, with specific adjacency relations and orientations. The general form of $Z_{\sigma}$ we consider in this paper is 
\begin{equation}
Z_\sigma=\int_{SU(2)}\! dh_{vf}\prod_v W_v(h_{vf})\;\prod_{f} \delta(\prod_{v\in f} h_{vf}),
\label{Zsigma}
\end{equation}
namely an \emph{holonomy} formulation of spin foams \cite{Bianchi:2010mw}, where the variables $h_{vf}$ are analogous to the canonical variables in \eqref{eq:gauge inv}. The partition function $Z_{\sigma}$ is \emph{local} in space-time, i.e. it is given by a product of elementary vertex amplitudes $W_v(h_{vl})$, and 'face amplitudes' which impose a local condition on holonomies. 

The ordered product inside the face amplitude is over a cyclic sequence of vertices, according to the face orientation. The internal holonomies $h_{vf}$ have a vertex label $v$ and a face label $f$, so that they can be uniquely associated to wedges\footnote{A wedge labeled by $vf$ is a portion of a face $f$ adjacent to the vertex $v$.}. Alternatively, the internal holonomies can be thought as associated to the links in the boundary graphs of vertices. We shall call these graphs \emph{local boundary} graphs, or simply local boundaries, in order to distinguish them from the \emph{global boundary} of the spin foam. The boundary graph of a vertex is defined as follows:  the links and the nodes of the boundary graph result from the intersection between the faces and the edges meeting at the vertex  with the boundary of a small 4-ball containing the vertex. The orientation of the boundary graph is inherited from the orientation of the 2-complex.

For spin foams with boundary graph $\Gamma$, the partition function becomes a function of boundary holonomies and generalizes to
\begin{align}
Z(h_l)=\sum_{\partial\sigma=\Gamma} W_\sigma(h_l),
\label{ZetaBound}
\end{align}
where the sum is over 2-complexes bounded by $\Gamma$ and the amplitude associated to each 2-complex is
\begin{align}\nonumber
W_\sigma(h_l)&=\int_{SU(2)}\! dh_{vf}\prod_v W_v(h_{vf})\times\\&\times\prod_{f_{int}} \delta(\prod_{v\in f} h_{vf})\prod_{f_{ext}} \delta(h_l\prod_{v\in f} h_{vf}).
\end{align}
To simplify the notation, the face amplitudes have been split in internal times external. The ones associated to external faces (faces cutting the boundary surface $\Sigma$ through a link $l$) contain the boundary holonomy $h_l$.
\begin{figure}[h!]
\includegraphics[width=4.5cm]{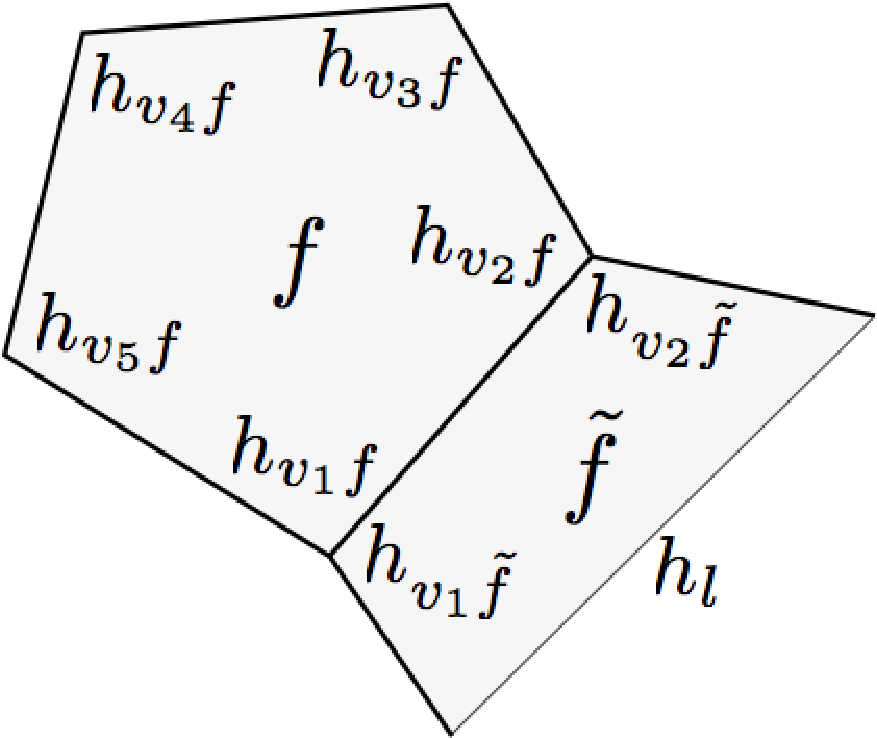}
\caption{Labeling of spin foam in the holonomy representation. An internal face $f$ and the external face $\tilde f$ are shown. The holonomies $h_{vf}$ are associated to wedges, and the holonomy $h_l$ to the external face $\tilde f$, or equivalently to the boundary link $l$.}
\label{fig:labeling}
\end{figure}
The amplitude $W_\sigma(h_l)$ defines a linear functional:
\begin{align}
W_\sigma: \mathcal H_\Gamma\longrightarrow\mathbbm C
\end{align}
where $\mathcal H_\Gamma=L^2(SU(2)^L)/\sim$ is the boundary Hilbert space associated to the boundary graph $\Gamma$ (we have divided by the gauge action of $SU(2)$ at nodes). This functional is
\begin{align}
(W_\sigma|\psi\rangle=\int_{SU(2)} dh_l W(h_l)\psi(h_l),
\end{align}
and assigns a quantum amplitude to the kinematical states. This amplitude (its modulus) gives the probability for a joint set of measurements, coded in the boundary state $\psi$ performed on the boundary of $\sigma$ \cite{Oeckl:2003vu,Rovelli:2004tv}.

The formalism admits boundaries with two connected components $\Sigma_{in}$ and $\Sigma_{out}$. In this case, $W_\sigma$ belongs to $\mathcal H^*_{\Gamma_{out}}\otimes \mathcal H_{\Gamma_{in}}$ and  can be thought as a propagation kernel evolving the incoming state $\psi_{in}$ to the outgoing state $\psi_{out}$
\begin{align}
\psi_{out}(h_{l'})=\int_{SU(2)} dg_{l} W(h_{l'},g_l)\psi_{in}(g_l)
\end{align}
so we recover the more standard interpretation of $W_{\sigma}$ as a transition amplitude.
\section{Local spin foams}
In this section we analyze the properties of a local spin foam of the form \eqref{Zsigma}, and argue that the requirement of cylindrical consistency can fix the arbitrary normalization of the vertex amplitude. At the conceptual level, cylindrical consistency is an important step for the interpretation of a spin foam as a Feynman path-integral over histories of the connection $A_\mu(\vec x,t)$. The type of consistencies we discuss in this section are 'canonical', in the following sense; they are a sort of time-evolution of the well-known cylindrical consistency for LQG operators (e.g. the volume operator \cite{Thiemann:2007zz}): link reversal, link splitting and link erasing are lifted to the spin foam framework.

In the following we shall omit the decoration of the integrals with the corresponding group. In general, the lower case variables $h$ and $g$ are $SU(2)$ holonomies, whereas the upper case variables $G$ are in the full 4-dimensianal gauge group $Spin(4)$ or $SL(2,\mbb C)$ depending on the space-time signature.
\begin{enumerate}
\item \emph{Locality.} As we already mentioned, the spin foam partition function \eqref{Zsigma} is local. Actually, formula \eqref{Zsigma} defines a notion of locality. The standard locality in terms of colorings of the 2-complex is recovered   
using the Peter-Weyl decomposition:
\begin{align}
Z_{\sigma}=\sum_{j_f i_e}\prod_f (2 j_f+1)\prod_e A_e(j_f,i_e)\prod_v A_v(j_f, i_e)
\label{ZPW}
\end{align}
Here the spins $j_f$ label the spin foam faces. Notice that the face amplitude is the dimension of the $SU(2)$ representation $d_j=2j+1$, for any spin foam model of the form \eqref{Zsigma}.

The edge and vertex amplitudes $A_e$ , $A_v$ are local: $A_e$ depends only on the intertwiner labeling the edge and the spins of the faces meeting at the edge, $A_v$ depends only on the the quantum numbers of edges and faces meeting at the vertex. Finally, the edge amplitude can be absorbed in a redefinition of the vertex amplitude. As a consequence, possible normalization ambiguities are absorbed in the vertex amplitude.
\item \emph{Composition property (face cutting).} The spin foam amplitude $W_{\sigma}(h_{vf})$ satisfies a composition property under face cutting, emphasized in \cite{Bianchi:2010fj}. Suppose we cut the 2-complex in two pieces $\sigma_1$ and $\sigma_2$ (the cut is realized by intersecting with a 3-surface), in such a way that each face which has been cut turns into two external faces of $\sigma_1$ and $\sigma_2$ respectively. We have
\begin{align}
\int dh_{ext} \overline{W_{\sigma_1}(h,h_{ext})} W_{\sigma_2}(h_{ext},h)=W_{\sigma_1\cup\sigma_2}(h)
\end{align}
where the integration is over the boundary holonomies $h_{ext}$. This follows easily from the following simple property of the face amplitudes
\begin{align}
\int d h_{ext}\delta(h\ldots h_{ext})\delta(h'\ldots h_{ext})=\delta(h\ldots h'\ldots),
\end{align}
namely the two external face amplitudes collapse into a single internal face amplitude, after integration. If we change the face amplitude in \eqref{ZPW}, the composition property does not hold anymore \cite{Bianchi:2010fj}. In particular, this fixes the face amplitude of EPRL model to be $d_j$, and \emph{not} the $SO(4)$ one $d_{j^+}d_{j^-}$, or even worse the $SL(2,\mathbbm C)$ dimension in the Lorentzian case, which is infinite.
\item \emph{Cilindrical consistency (face reversal).}
As a first step, we require face reversal invariance of the spin foam model (Fig.\ref{fig:facereversal}). 
\begin{figure}[h!]
\parbox{2cm}{\includegraphics[width=2cm]{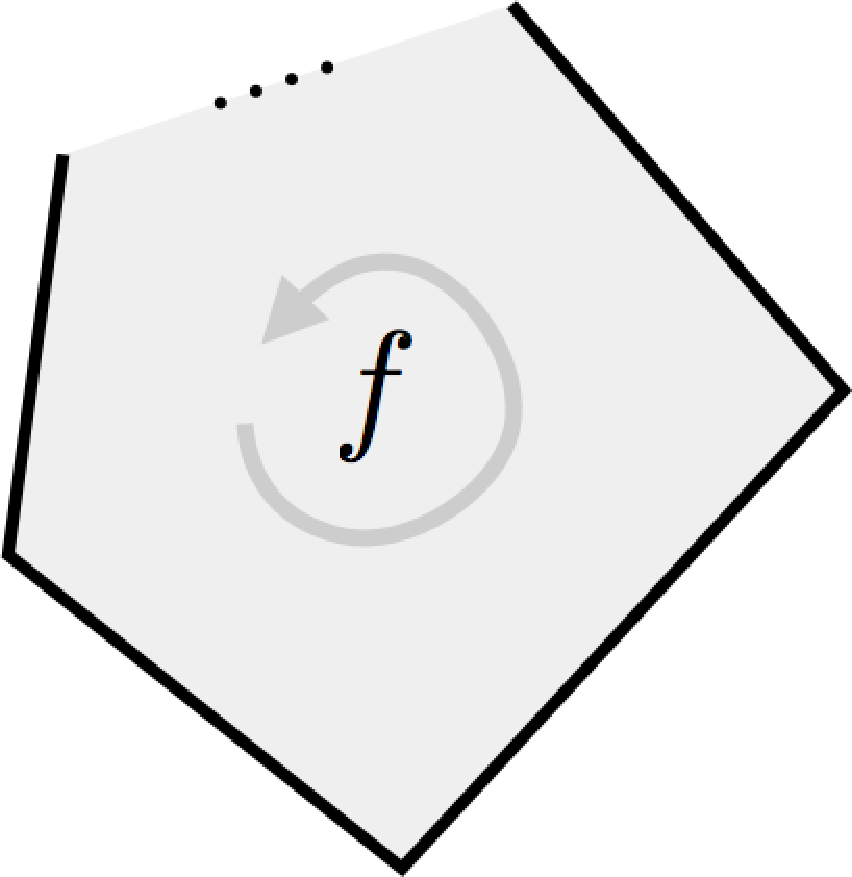}}\parbox{1.5cm}{$\rightarrow$}\parbox{2cm}{\includegraphics[width=2cm]{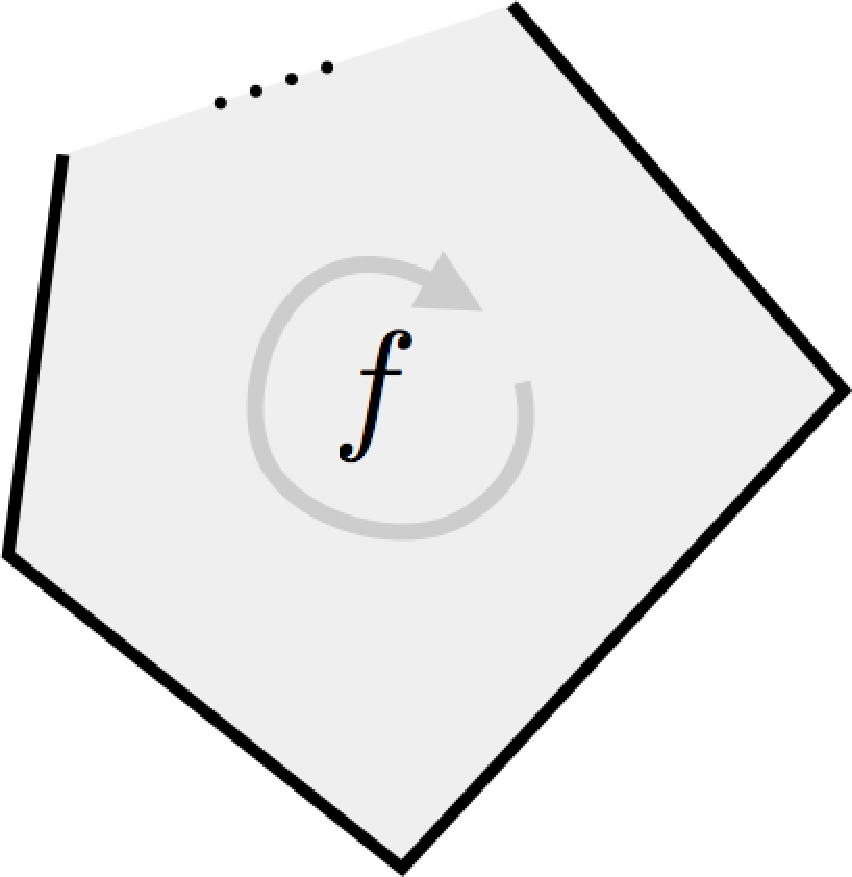}}
\caption{The face reversal.}
\label{fig:facereversal}
\end{figure}
Under a flip in the orientation of a face $f$, the face amplitude has the transformation rule
\begin{align}
\delta(\prod_{v\in f}h_{vf})\rightarrow \delta(\prod_{v\in f}h^{-1}_{vf})
\end{align}
where the two products have the \emph{same} ordering. 
It follows that the partition function is invariant if we assume the following transformation rule for the vertex amplitudes of vertices belonging to the same face $f$:
\begin{align}
W_{v}(h_{vf},\ldots)\rightarrow W_v(h^{-1}_{vf},\ldots)
\label{reversalvertex}
\end{align}
\item \emph{Cilindrical consistency (face splitting).} 
Consider a holonomy $h_l$ associated to a link in the local boundary of a vertex, or in the global boundary of a 2-complex. If we split the link in two parts $l'\cup l''=l$ and associate to each part a holonomy, we would like to regard the product $h_{l'} h_{l''}$ as equivalent to the single holonomy $h_l$ (Fig.\ref{fig:linksplitting}). 
\begin{figure}[h!]
\includegraphics[width=3.5cm]{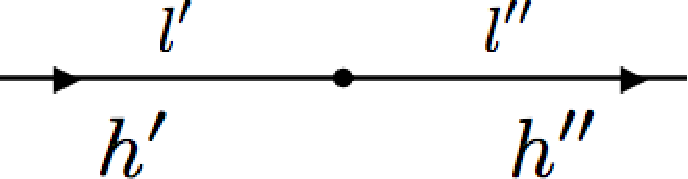}
\caption{The composition low of holonomies. $H_{l=l'\cup l''}(A)=H_{l'}(A)H_{l''}(A)$.}
\label{fig:linksplitting}
\end{figure}
This picture comes from the composition law of holonomies in the connection representation of Loop Quantum Gravity, where the holonomy of a connection $A_\mu$ satisfies
\begin{align}
H_{l=l'\cup l''}(A)=H_{l'}(A)H_{l''}(A)
\end{align}
if the path $l''$ starts where the path $l'$ ends.

The fact that in the spin foam picture the two objects are different has important consequences for the dynamics. Consider an $N$-valent spin foam vertex and its local boundary whose links are labeled by $vf$. The vertex amplitude is $W_v(h_{vf})$. Let us split one of the boundary links in two pieces, with associated holonomies $h_{vf'}$ and $h_{vf''}$. Since we added a dummy (i.e. 2-valent) node, we have split the face $f$ in two parts (Fig.\ref{fig:splitting})
\begin{figure}[h!]
\includegraphics[width=4.5cm]{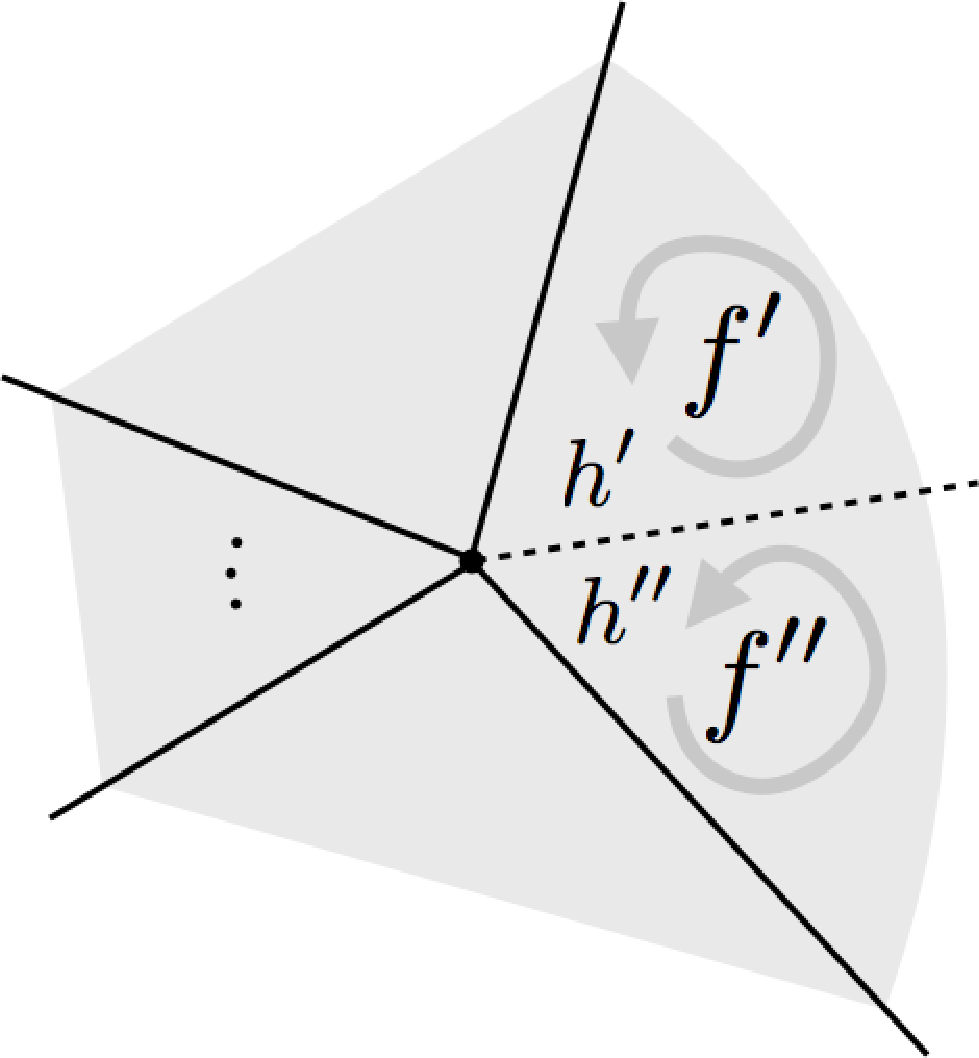}
\caption{The face splitting. A face $f$ is split in two faces $f'$, $f''$ by a new dummy (2-valent) edge. The holonomy labeling the original face is split in two independent holonomies $h'$, $h''$.}
\label{fig:splitting}
\end{figure}
by adding a dummy (2-valent) edge that comes out from the vertex. Now we have actually increased the valence of the vertex by 1. The new vertex amplitude reads $W_v(h_{vf'},h_{vf''},\ldots)$. In general the two vertex amplitudes (the $N$ and the $(N+1)$-valent) can be different, so we require:
\begin{align}
W_v(h_{vf'},h_{vf''},\ldots)= W_v(h_{vf'}h_{vf''},\ldots).
\label{splitrequirement}
\end{align}
This requirement provides a cylindrical consistency for spin foam theory: if we use redundant holonomies in the description of the kinematical space, the quantum amplitudes should not depend on this choice. Physics must be independent of it. It is important to notice that the splitting edge can end in another vertex or in the (global) boundary, if there is one. It is straightforward to show that for a general model \eqref{Zsigma}, equation \eqref{splitrequirement} implies that the partition function is invariant under this splitting. More precisely, $W_\sigma$ is invariant under a vertex-to-vertex splitting, and satisfies
\begin{align}
W_{\tilde\sigma}(h_{l'},h_{l''},\ldots)= W_\sigma(h_{l'}h_{l''},\ldots)
\label{splittoboundary}
\end{align}
for a splitting edge that ends on the boundary (Fig.\ref{fig:splittingtotal} gives a pictorial representation of the two kinds of splitting).
\begin{figure}[h!]
\includegraphics[width=3.3cm]{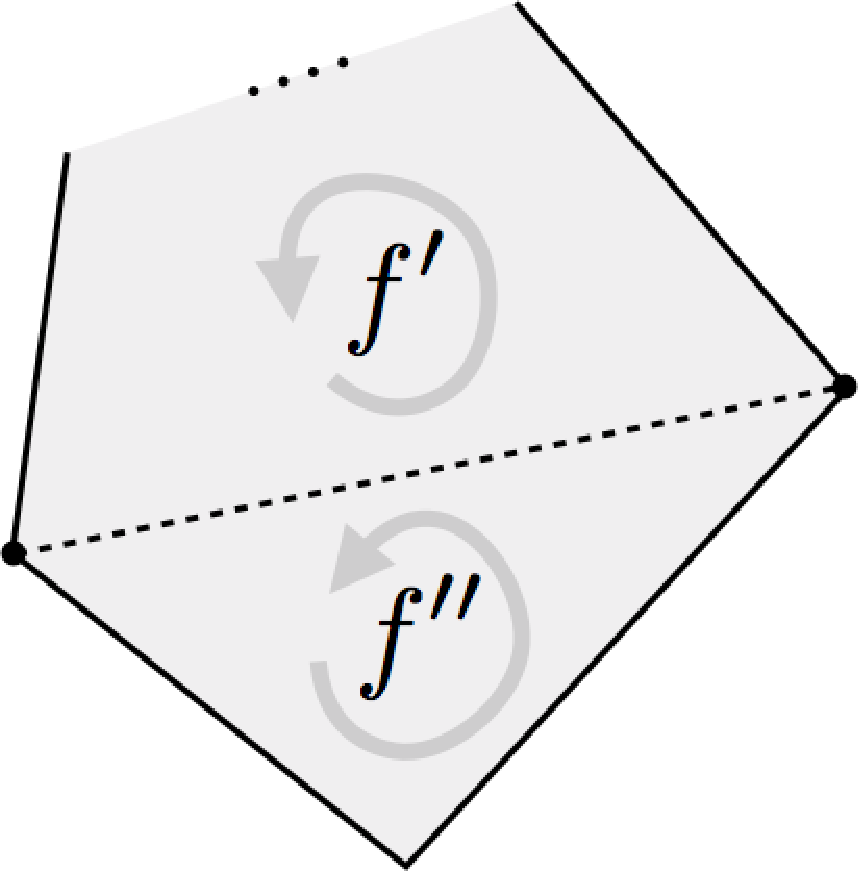}\quad\quad\quad
\includegraphics[width=2.85cm]{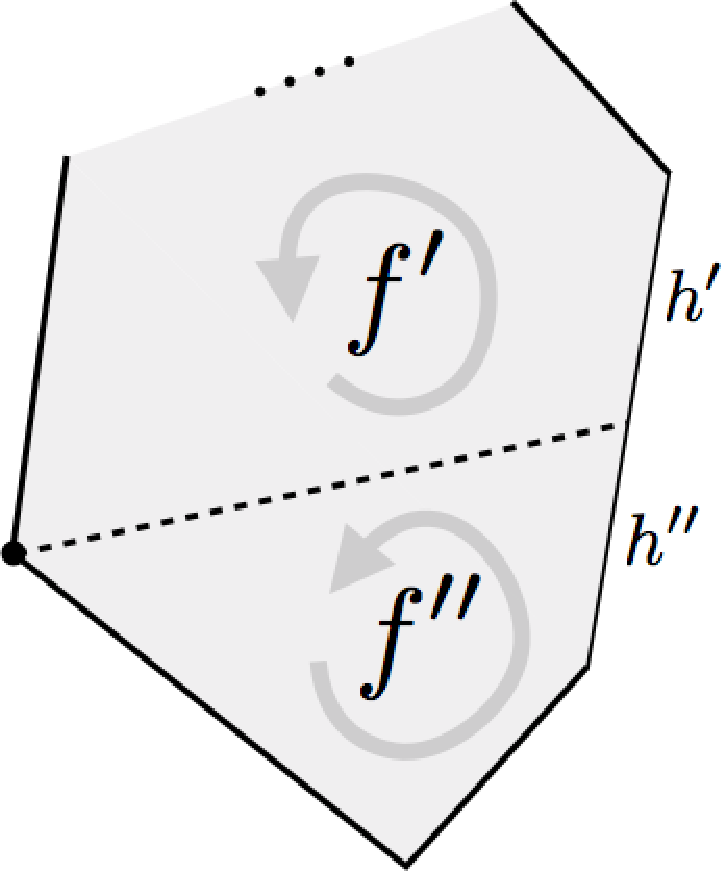}
\caption{On the left: the vertex-to-vertex face splitting. A face $f$ is split in two faces $f'$, $f''$ by a new dummy (2-valent) edge. On the right: the face splitting ending on the boundary.}
\label{fig:splittingtotal}
\end{figure}
The cylindrical requirement \eqref{splitrequirement}  is trivially met for the Ponzano-Regge model, while in the case of EPRL it fixes the residual ambiguity in the definition of vertex amplitude. This is done in the following two subsections.
\item{\emph{Cylindrical consistency (face erasing)}}
As a last requirement, we demand the spin foam amplitude $W_\sigma$ for a 2-complex $\sigma$ bounded by the graph $\Gamma$  to satisfy:
\begin{align}
(W_{\sigma}|\psi_{\Gamma,j}\rangle=(W_{\tilde{\sigma}}|\psi_{\tilde\Gamma,j}\rangle
\label{spinerasing}
\end{align}
where $\tilde\sigma$ is obtained from $\sigma$ by erasing some of the external faces, the new boundary $\tilde\Gamma$ is the subgraph of $\Gamma$ obtained by erasing the corresponding links, and the spins of the spin-network $\psi_{\Gamma,j}$ labeling $\Gamma-\tilde\Gamma$ have been set to zero (Fig.\ref{fig:faceerasing}).
\begin{figure}[h!]
\includegraphics[width=3.5cm]{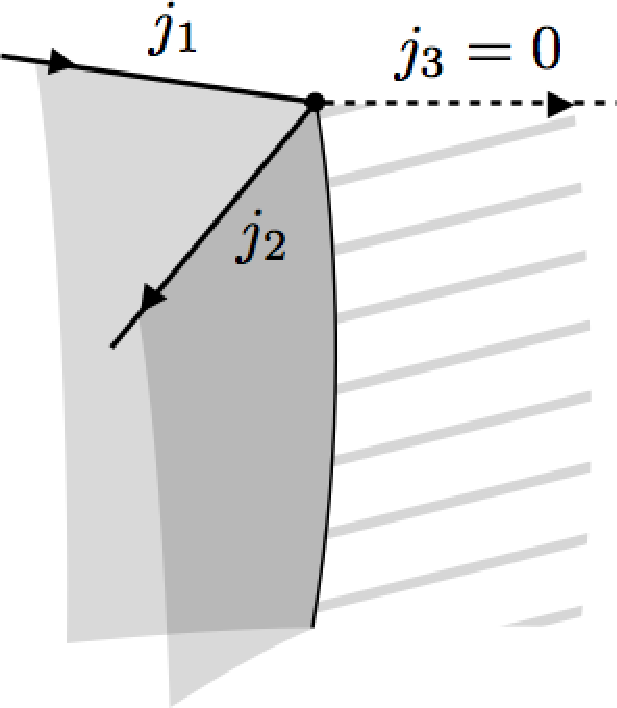}
\caption{The face erasing.}
\label{fig:faceerasing}
\end{figure}
The cylindrical consistency requirement \eqref{spinerasing} can be translated in a requirement on the colorings of the 2-complex.

In fact, \eqref{spinerasing} implies the following thing. Consider the Peter-Weyl expansion of $Z_\sigma$ over colorings of the 2-complex, as in \eqref{ZPW}, and consider a generic term of this sum such that some of the spins vanish. Then this term will coincide with an analogous term in the partition function $Z_{\tilde\sigma}$, where the sub-foam $\tilde\sigma\subset\sigma$ is derived from $\sigma$ by erasing all the faces labeled by vanishing spins and glueing faces along trivial (2-valent) edges.

Consistency under face erasing is strictly related to consistency under face splitting. Indeed even if we start with a $n$-valent edge ($n>2$) we can erase $n-2$ faces and end up with a trivial edge (see Fig.\ref{fig:faceerasing}). This situation is clearly equivalent to a face splitting, so property \eqref{splitrequirement} ensure that the trivial edge can be safely removed.
\end{enumerate}
\subsection*{Ponzano-Regge model}
The Ponzano-Regge model for three-dimensional Euclidean quantum gravity \cite{Regge:1961px,PonzanoRegge:1968,Boulatov:1992vp} is defined by the general formula \eqref{Zsigma} with the following vertex amplitude
\begin{equation}
W_v^{PR}(h_{vf})=\int d g_{ve}\; \prod_{vf} \delta(h_{vf},\, g^{-1}_{t(vf)}g_{s(vf)})
\label{eq:PR h}
\end{equation}
where the integration is over gauge group variables $g_{ve}$ (one per each edge $e$ coming out from the vertex $v$), and $s$, $t$ specify if the edge is 'source' or 'target' according to the orientation of faces. Clearly, the interpretation of variables is slightly different from the four-dimensional models. Here the Ponzano-Regge model is viewed as a covariant path-integral formulation of 2+1 Loop Quantum Gravity \cite{Noui:2004iy}. Therefore the $SU(2)$ variables $h_{vf}$ are interpreted as holonomies of the $SU(2)$ spin-connection $\omega_\mu$. Moreover, local boundaries of spin foam vertices are defined by intersection with a small 3-ball. In the following we review well-known invariance properties of the Ponzano-Regge model in the language of the new holonomy formulation for local spin foams. The proofs will be similar in the four-dimensional model, therefore we consider 2+1 gravity as a warm-up.

As a first step, we show that the Ponzano-Regge partition function is left invariant under face reversal. To achieve this, it suffices to prove the transformation rule of vertex amplitude \eqref{reversalvertex}. So consider a vertex $v$ belonging to the face $f$, and consider the face factor ($\delta$ distribution) in the vertex amplitude associated to this face. Under a flip in the orientation of $f$, the source and target gauge variables must be interchanged. But from the simple identity $\delta(g)=\delta(g^{-1})$ we have
\begin{align}
\delta(h_{vf},\,g_{s(vf)}^{-1}g_{t(vf)})=\delta(h_{vf}^{-1}, \,g^{-1}_{t(vf)}g_{s(vf)})
\label{PRflipproof}
\end{align}
hence we prove \eqref{reversalvertex}.

The second cylindrical consistency we consider is face splitting invariance, in particular we prove \eqref{splitrequirement}. Consider a vertex $v$ of a face $f$ and let us split $f$ in two new faces $f'$ and $f''$. To simplify the notation, call $g_{s(vf')}=g_{t(vf'')}=g$ the gauge associated to the 2-valent edge where $f'$ and $f''$ meet. Call also $h_{vf'}$ and $h_{vf''}$ simply $h'$ and $h''$ respectively, $g_{t(vf')}=t$ and $g_{s(vf'')}=s$. Then the new vertex will contain the two face factors
\begin{align}
\delta(h',t^{-1} g)\delta(h'',g^{-1} s).
\label{twofactors}
\end{align}
Performing the integral in $g$, we have
\begin{align}
\int dg\,\delta(h',t^{-1} g)\delta(h'',g^{-1} s)=\delta(h' h'', t^{-1} s)
\label{PRsplitproof}
\end{align}
so the two face factors \eqref{twofactors} collapse into a single face factor. This shows that the requirement \eqref{splitrequirement} holds.

As a last step, we prove cylindrical consistency under face erasing. In particular, we prove \eqref{spinerasing} for a single vertex, the generalization being straightforward. A normalized spin-network\footnote{We have omitted the intertwiner labels in the notation, since they play no role in our analysis.} function $\psi_{\Gamma j}$ on the local boundary graph $\Gamma$ has the property of being equal to the spin-network $\psi_{\tilde\Gamma j}$ on a sub-graph $\tilde\Gamma\subset\Gamma$ when the spins of $\Gamma-\tilde\Gamma$ are set to zero. The cylindrical consistency of the spin foam model requires the corresponding vertex amplitudes to be equal. This is true, due to the following simple observation: when evaluating $(W_v|\psi_j\rangle$, some of the face factors in the vertex amplitude are integrated against trivial (spin zero) Wigner matrices, and give trivial contribution to the evaluation:
\begin{align}
\int dh_{vf} \,\delta(h_{vf},\, g^{-1}_{t(vf)}g_{s(vf)})D^{j_{vf}=0}(h_{vf})=1.
\label{PRerasing}
\end{align}
\subsection*{EPRL model} The EPRL model \cite{Engle:2007wy} is a candidate model for quantum general relativity in four dimensions. Here we are interested in a manifestly $SU(2)$-invariant holonomy formulation of this model, which appeared in the local form \eqref{Zsigma} in reference \cite{Bianchi:2010mw}. The model is specified by the following vertex amplitude:
\begin{equation}
W_v^{EPRL}(h_{vf})=\int dG_{ve} \; \prod_{vf} \delta_\gamma(h_{vf},\, G^{-1}_{t(vf)}G_{s(vf)})
\label{EPRLvertex}
\end{equation}
where the integral is over the $Spin(4)$ ($SL(2,\mathbbm C)$ in the Lorentzian theory) gauge group variables $G_{ve}$ and the source/target group elements are defined according to the orientation of faces. We have introduced the following distribution:
\begin{align}
\delta_\gamma(h,G)=\begin{cases}\sum_j \int_{SU(2)} dc\,N_j \,\chi^j(c \,h)\chi^{(j^+,j^-)}(c \,G)\quad\quad(e)\vspace{0.2cm}
\\\sum_j \int_{SU(2)} dc\,N_j \,\chi^j(c \,h)\chi^{(n,\rho)}(c \,G)\quad\quad\quad(l)
\end{cases}
\label{deltagamma}
\end{align}
where the two lines refer to the Euclidean (e) and Lorentzian (l) versions of the model respectively, and $\gamma\neq 1$ is the Barbero-Immirzi parameter. The irreducible representations labeling the characters $\chi$ of $SO(4)$ (or $SL(2,\mathbbm C)$)  in \eqref{deltagamma} satisfy the simplicity constraint:
\begin{align}
\begin{cases}
j^+=\frac{1+\gamma}{2}j,\quad j^-=\frac{|1-\gamma|}{2}j\;\quad\quad(e)\\
n=2j,\quad \rho=2\gamma j\;\,\,\;\quad\quad\quad\quad\quad(l)
\end{cases}
\end{align}
The simple form (\ref{EPRLvertex}, \ref{deltagamma}) of the vertex is derived from the one of reference \cite{Bianchi:2010mw}. Basically, the embedding maps $SU(2)\rightarrow SO(4)$ (or $SU(2)\rightarrow SL(2,\mathbbm C)$) are replaced by an auxiliary integration over the $c$ variable ($SU(2)$-averaging), which is responsible for the coupling between the little group $SU(2)$ with the 4-dimensional gauge group. The positive constant $N_j$ in \eqref{deltagamma} parametrizes the normalization of the vertex (and edge) amplitude. In the Euclidean model, the vertex normalization $N_j$ can be fixed from the requirement of cylindrical consistency. This could be done in principle also in the Lorentzian theory, but we need to handle potential divergencies resulting from the non-compactness of $SL(2,\mathbbm C)$, and will be discussed elsewhere.

Following the three steps as for the Ponzano-Regge model, we first study the effect of a flip in the orientation of a face. The identity we have to prove, analogous to \eqref{PRflipproof} is the following:
\begin{align}
\delta_\gamma(h_{vf},\, G_{s(vf)}^{-1}G_{t(vf)})=\delta_\gamma(h_{vf}^{-1},\, G^{-1}_{t(vf)}G_{s(vf)}).
\end{align}
This is easily done using the formalism \eqref{deltagamma}:
\begin{align}
&\delta_\gamma(h,G^{-1})=\nonumber\\&=\sum_j \int dc\,N_j \chi^j(c g) \chi^{j^+}(c (g^+)^{-1}) \chi^{j^-}(c (g^-)^{-1})=\nonumber\\
&=\sum_j \int dc\,N_j \chi^j(c^{-1} g^{-1}) \chi^{j^+}(c^{-1} g^+) \chi^{j^-}(c^{-1} g^-)=\nonumber\\
&=\delta_\gamma(h^{-1},G)
\end{align}
where we used the simple property $\delta(g)=\delta(g^{-1})$ and the cyclic invariance of traces.

Our second goal is to prove the face splitting rule \eqref{splitrequirement}. After a face splitting (as in \eqref{twofactors}), one face factor in the corresponding vertex splits into two parts
\begin{align}
\delta_\gamma(h',T^{-1} G)\delta_\gamma(h'',G^{-1} S).
\label{twofactorsEPRL}
\end{align}
Next, integration of the previous formula over $G\in SO(4)$ gives
\begin{align}
\sum_j\int_{SU(2)} dc\,\frac{N_j^2}{d_j d_{j^+} d_{j^-}}\,\chi^j(c h'h'')\chi^{(j^+,j^-)}(c T^{-1}S).
\end{align}
Therefore, choosing the normalization\footnote{The same normalization is considered in \cite{Geloun:2010vj}, for the Group Field Theory formulation.}
\begin{align}
N_j=d_j d_{j^+} d_{j^-},
\label{EPRLnorm}
\end{align}
expression \eqref{twofactorsEPRL} collapses to the single $\delta_\gamma$ distribution
\begin{align}
\delta_\gamma(h'h'',T^{-1} S).
\end{align}

As a final step, we prove the cylindrical consistency of EPRL model under face erasing. Here the analogous of equation \eqref{PRerasing} is, after a simple calculation,
\begin{align}
\int dh_{vf} \,\delta_\gamma(h_{vf},\, G^{-1}_{t(vf)}G_{s(vf)}) D^{j_{vf}=0}(h)=N_{j=0}.
\label{EPRLerasing}
\end{align}
Therefore, imposing the face splitting invariance, we have $N_{j=0}=1$ and prove the full cylindrical consistency of EPRL model with vertex normalization \eqref{EPRLnorm}.

Before concluding, we observe that cylindrical consistency of EPRL model allows to extend to the full Loop Quantum Gravity kinematical space \eqref{fullLQG} the embedding map $f_\gamma$ of $SU(2)$ spin-networks into (simple) $SO(4)$ spin-networks. This map is the key ingredient for the definition of the model \cite{Engle:2007wy}. With our notations, the embedding map 
\begin{align}
f^{\Gamma}_\gamma: \mathcal H_\Gamma\rightarrow \mathcal H^{SO(4)}_\Gamma,
\end{align}
with $\mathcal H^{SO(4)}_\Gamma=L^2(SO(4)^L/\sim)$, is defined as
\begin{align}
(f^{\Gamma}_\gamma \psi)(H_l)=\int dG_ndh_l\,\prod_l\delta_\gamma(h_l,G^{-1}_{t(l)}H_l G_{s(l)})\psi(h_l).
\end{align}
To our knowledge, this map has been defined only for an arbitrary, but \emph{fixed}, graph. In order to define it consistently on all graphs, we need to check its cylindrical properties. Basically, the action on the space $\mathcal H_\Gamma$ have to be consistent with the action on $\mathcal H_{\tilde\Gamma}$, where $\Gamma$ is a sub-graph $\tilde\Gamma\subset\Gamma$. A sub-graph can be obtained from a larger graph with a finite number of elementary operations, which constist in flipping the orientation of links, splitting a link, or erasing links. It is not hard to understand that the three cylindrical requirements discussed in this paper imply that (at least with the normalization \eqref{EPRLnorm}) the embedding map $f_\gamma$ extends to the full Hilbert space
\begin{align}
f_\gamma: L^2(\overline{\mathcal A},d\mu_{AL})\rightarrow L^2(\overline{\mathcal A}^{SO(4)},d\mu^{SO(4)}_{AL})
\end{align}
where
\begin{align}
L^2(\overline{\mathcal A}^{SO(4)},d\mu^{SO(4)}_{AL})\simeq L^2(\overline{\mathcal A}_+\times \overline{\mathcal A}_-,d\mu^+_{AL}d\mu^-_{AL})
\end{align}
is built up from two copies (the 'left' one denoted by $+$, the 'right' one by a $-$) of the kinematical state space of loop gravity. Clearly, a basis for the image of $f_\gamma$ is given by simple spin-networks (see \cite{Dupuis:2010jn} for a recent review and analysis).
\section{Taking the continuum limit}
Consider the full partition function
\begin{align}
Z(h_l)=\sum_\sigma W_\sigma(h_l)
\label{Zbound2}
\end{align}
with sum over 2-complexes which are bounded by a graph $\Gamma$. The physical intuition suggests that a given 2-complex acts as a regulating lattice. In fact, spin foam models are generally defined as a quantization of a truncation of classical General Relativity to a finite number of degrees of freedom, e.g. by first discretizing it over a piecewise flat simplicial manifold, then quantizing it \cite{Ooguri:1992eb,Barbieri:1997ks,Barrett:1997gw,Engle:2007wy}. Differently from a lattice gauge theory, expression \eqref{Zbound2} gets rid of the regulator dependence by summing over all possible 2-complexes. This sum, including possible symmetry factors, can be generated by a Group Field Theory \cite{Oriti:2006se,Oriti:2009wn}.

However, a closer look at formula \eqref{Zbound2} suggests that there is a large amount of redundancy in this sum. Let us expand \eqref{Zbound2} in a sum over \emph{colored} 2-complexes:
\begin{align}
Z(h_l)=\sum_\sigma \sum_{j,i} W_{\sigma,j,i}(h_l).
\label{Zcolored}
\end{align}
A term $W_{\sigma,j,i}$ where some of the $j$'s are vanishing can be naturally interpreted as the amplitude for a sub 2-complex $\tilde \sigma$ obtained by erasing the corresponding faces. Since we are summing over the sub 2-complexes this term is counted at least twice. The degeneracy is clearly proportional to the number of vanishing spins in the most refined 2-complex $\sigma$. 
To avoid this overcounting, the sum over colorings \eqref{Zcolored} should be restricted to nonvanishing spins.

A different way for recovering the infinite number of degrees of freedom of General Relativity is to consider the partition function for a very fine 2-complex and take the limit of infinite refinement:
\begin{align}
Z(h_l)=\lim_{\sigma\rightarrow\sigma_\infty} W_\sigma(h_l).
\label{ZlatticeLIM}
\end{align}
This approach is much more similar to lattice gauge theories or dynamical triangulations \cite{Ambjorn:1992aw,Oeckl:2002ia}. Despite the two partition functions \eqref{Zbound2} and \eqref{ZlatticeLIM} look very different, they are likely to be related, or even identical. This observation was pointed out recently by Rovelli \cite{Rovelli:2010vv,RovSmer}. 

An insight on the the relation between the refinement limit \eqref{ZlatticeLIM} and a sum over 2-complexes comes from the consistency requirements discussed in this paper. Let us put a cut-off on the theory, namely consider a very fine 2-complex $\sigma$. The quantity
\begin{align}
Z_{(\sigma)}(h_l)=W_\sigma(h_l)
\label{Zcutoff}
\end{align}
can be interpreted as a cut-off 'a la lattice gauge theory' of the partition function. An alternative definition of partition function as a sum over 2-complexes with the cut-off induced by the choice of $\sigma$ is
\begin{align}
Z'_{(\sigma)}(h_l)=\sum_{\rho\subset\sigma}\sum_{j\neq 0,i}W_{\rho,j,i}(h_l)
\label{Zpcutoff}
\end{align}
where the sum is over sub 2-complexes and over non-trivial colorings, and the sub 2-complexes $\rho$ in \eqref{Zpcutoff} are the ones that can be obtained from $\sigma$ by erasing faces in all possible ways, and possibly, eliminating the trivial edges. Notice that in order to merge two faces along a trivial edge, we have to ensure that they have compatible orientations (they must induce opposite orientations on the common edge).

Hence the sub-foams $\rho$ are obtained via a finite number of elementary operations. These are the face orientation reversal, the face splitting, and the face erasing. So if the cylindrical requirements discussed in this paper are satisfied, as for the EPRL spin foam model, the partition functions \eqref{Zcutoff} and \eqref{Zpcutoff} are the same.
 
More difficult is to show that a sum like \eqref{Zpcutoff} corresponds to a sum over equivalence classes of 2-complexes similarly to Group Field Theories (see \cite{Zapata:2002eu} for a complete analysis in 2+1 gravity). To this regard a possible difficulty could come from trivial vertices, that is vertices of valence two, bounding at least three faces. Preliminary investigations indicate that, in the case of EPRL model (at least its $SU(2)$ version) the trivial vertices cannot be erased without affecting the partition function. We leave this as an open problem. 
\section{Conclusions}
In this paper a new holonomy formulation for spin foams was shown to be an appropriate tool to deal with general features of spin foam models. Within the holonomy representation, we introduced cylindrical consistency for spin foams as a natural step towards a contact with the (Ashtekar) connection representation of canonical Loop Quantum Gravity. We discussed an important consequence of cylindrical consistency: it fixes the arbitrary normalization of the vertex amplitude of EPRL model. Furthermore, it provides key insights on the continuum limit. The extension of our analysis to the Lorentzian signature is left for a future investigation. In the Lorentzian case, the volume of the gauge group is infinite, and so is the dimension of the irreducible unitary representations. This could affect the cylindrical consistency condition (face splitting) we have introduced.

\begin{acknowledgments}
The idea of implementing cylindrical consistency in the 'new' spin foam models came some time ago in Marseille. At that time we lacked a simple formalism (the local holonomy formulation) to handle this problem. Our interest was renewed by a stimulating exchange of ideas with Jerzy Lewandowski, who gave a series of three lectures ``Canonical Loop Quantum Gravity and Spin Foams'' in Penn State, during August 2010. A warm thank goes to Carlo Rovelli, Matteo Smerlak and Antonino Marcian\`o for useful discussions and comments on the manuscript. This work was supported in part by the NSF grant PHY0854743, The George A.\ and Margaret M.~Downsbrough Endowment and the Eberly research funds of Penn State. E.M. gratefully acknowledges support from Fondazione A. della Riccia.
\end{acknowledgments}
\bibliography{biblioLSF}

\begin{thebibliography}{49}%
\makeatletter
\providecommand \@ifxundefined [1]{%
 \@ifx{#1\undefined}
}%
\providecommand \@ifnum [1]{%
 \ifnum #1\expandafter \@firstoftwo
 \else \expandafter \@secondoftwo
 \fi
}%
\providecommand \@ifx [1]{%
 \ifx #1\expandafter \@firstoftwo
 \else \expandafter \@secondoftwo
 \fi
}%
\providecommand \natexlab [1]{#1}%
\providecommand \enquote  [1]{``#1''}%
\providecommand \bibnamefont  [1]{#1}%
\providecommand \bibfnamefont [1]{#1}%
\providecommand \citenamefont [1]{#1}%
\providecommand \href@noop [0]{\@secondoftwo}%
\providecommand \href [0]{\begingroup \@sanitize@url \@href}%
\providecommand \@href[1]{\@@startlink{#1}\@@href}%
\providecommand \@@href[1]{\endgroup#1\@@endlink}%
\providecommand \@sanitize@url [0]{\catcode `\\12\catcode `\$12\catcode
  `\&12\catcode `\#12\catcode `\^12\catcode `\_12\catcode `\%12\relax}%
\providecommand \@@startlink[1]{}%
\providecommand \@@endlink[0]{}%
\providecommand \url  [0]{\begingroup\@sanitize@url \@url }%
\providecommand \@url [1]{\endgroup\@href {#1}{\urlprefix }}%
\providecommand \urlprefix  [0]{URL }%
\providecommand \Eprint [0]{\href }%
\providecommand \doibase [0]{http://dx.doi.org/}%
\providecommand \selectlanguage [0]{\@gobble}%
\providecommand \bibinfo  [0]{\@secondoftwo}%
\providecommand \bibfield  [0]{\@secondoftwo}%
\providecommand \translation [1]{[#1]}%
\providecommand \BibitemOpen [0]{}%
\providecommand \bibitemStop [0]{}%
\providecommand \bibitemNoStop [0]{.\EOS\space}%
\providecommand \EOS [0]{\spacefactor3000\relax}%
\providecommand \BibitemShut  [1]{\csname bibitem#1\endcsname}%
\let\auto@bib@innerbib\@empty
\bibitem [{\citenamefont {Bianchi}\ \emph
  {et~al.}(2010{\natexlab{a}})\citenamefont {Bianchi}, \citenamefont
  {Magliaro},\ and\ \citenamefont {Perini}}]{Bianchi:2010mw}%
  \BibitemOpen
  \bibfield  {author} {\bibinfo {author} {\bibfnamefont {Eugenio}\ \bibnamefont
  {Bianchi}}, \bibinfo {author} {\bibfnamefont {Elena}\ \bibnamefont
  {Magliaro}}, \ and\ \bibinfo {author} {\bibfnamefont {Claudio}\ \bibnamefont
  {Perini}},\ }\bibfield  {title} {\enquote {\bibinfo {title} {{Spinfoams in
  the holomorphic representation}},}\ }\href@noop {} {\  (\bibinfo {year}
  {2010}{\natexlab{a}})},\ \Eprint {http://arxiv.org/abs/1004.4550}
  {arXiv:1004.4550 [gr-qc]} \BibitemShut {NoStop}%
\bibitem [{\citenamefont {Reisenberger}\ and\ \citenamefont
  {Rovelli}(2000)}]{Reisenberger:2000fy}%
  \BibitemOpen
  \bibfield  {author} {\bibinfo {author} {\bibfnamefont {Michael}\ \bibnamefont
  {Reisenberger}}\ and\ \bibinfo {author} {\bibfnamefont {Carlo}\ \bibnamefont
  {Rovelli}},\ }\bibfield  {title} {\enquote {\bibinfo {title} {{Spin foams as
  Feynman diagrams}},}\ }\href@noop {} {\  (\bibinfo {year} {2000})},\ \Eprint
  {http://arxiv.org/abs/gr-qc/0002083} {arXiv:gr-qc/0002083} \BibitemShut
  {NoStop}%
\bibitem [{\citenamefont {Perez}(2003)}]{Perez:2003vx}%
  \BibitemOpen
  \bibfield  {author} {\bibinfo {author} {\bibfnamefont {Alejandro}\
  \bibnamefont {Perez}},\ }\bibfield  {title} {\enquote {\bibinfo {title}
  {{Spin foam models for quantum gravity}},}\ }\href@noop {} {\bibfield
  {journal} {\bibinfo  {journal} {Class. Quant. Grav.}\ }\textbf {\bibinfo
  {volume} {20}},\ \bibinfo {pages} {R43} (\bibinfo {year} {2003})},\ \Eprint
  {http://arxiv.org/abs/gr-qc/0301113} {arXiv:gr-qc/0301113} \BibitemShut
  {NoStop}%
\bibitem [{\citenamefont {Livine}\ and\ \citenamefont
  {Speziale}(2007)}]{Livine:2007vk}%
  \BibitemOpen
  \bibfield  {author} {\bibinfo {author} {\bibfnamefont {Etera~R.}\
  \bibnamefont {Livine}}\ and\ \bibinfo {author} {\bibfnamefont {Simone}\
  \bibnamefont {Speziale}},\ }\bibfield  {title} {\enquote {\bibinfo {title}
  {{A new spinfoam vertex for quantum gravity}},}\ }\href {\doibase
  10.1103/PhysRevD.76.084028} {\bibfield  {journal} {\bibinfo  {journal} {Phys.
  Rev.}\ }\textbf {\bibinfo {volume} {D76}},\ \bibinfo {pages} {084028}
  (\bibinfo {year} {2007})},\ \Eprint {http://arxiv.org/abs/0705.0674}
  {arXiv:0705.0674 [gr-qc]} \BibitemShut {NoStop}%
\bibitem [{\citenamefont {Engle}\ \emph {et~al.}(2008)\citenamefont {Engle},
  \citenamefont {Livine}, \citenamefont {Pereira},\ and\ \citenamefont
  {Rovelli}}]{Engle:2007wy}%
  \BibitemOpen
  \bibfield  {author} {\bibinfo {author} {\bibfnamefont {Jonathan}\
  \bibnamefont {Engle}}, \bibinfo {author} {\bibfnamefont {Etera}\ \bibnamefont
  {Livine}}, \bibinfo {author} {\bibfnamefont {Roberto}\ \bibnamefont
  {Pereira}}, \ and\ \bibinfo {author} {\bibfnamefont {Carlo}\ \bibnamefont
  {Rovelli}},\ }\bibfield  {title} {\enquote {\bibinfo {title} {{LQG vertex
  with finite Immirzi parameter}},}\ }\href {\doibase
  10.1016/j.nuclphysb.2008.02.018} {\bibfield  {journal} {\bibinfo  {journal}
  {Nucl. Phys.}\ }\textbf {\bibinfo {volume} {B799}},\ \bibinfo {pages}
  {136--149} (\bibinfo {year} {2008})},\ \Eprint
  {http://arxiv.org/abs/0711.0146} {arXiv:0711.0146 [gr-qc]} \BibitemShut
  {NoStop}%
\bibitem [{\citenamefont {Freidel}\ and\ \citenamefont
  {Krasnov}(2008)}]{Freidel:2007py}%
  \BibitemOpen
  \bibfield  {author} {\bibinfo {author} {\bibfnamefont {Laurent}\ \bibnamefont
  {Freidel}}\ and\ \bibinfo {author} {\bibfnamefont {Kirill}\ \bibnamefont
  {Krasnov}},\ }\bibfield  {title} {\enquote {\bibinfo {title} {{A New Spin
  Foam Model for 4d Gravity}},}\ }\href {\doibase
  10.1088/0264-9381/25/12/125018} {\bibfield  {journal} {\bibinfo  {journal}
  {Class. Quant. Grav.}\ }\textbf {\bibinfo {volume} {25}},\ \bibinfo {pages}
  {125018} (\bibinfo {year} {2008})},\ \Eprint {http://arxiv.org/abs/0708.1595}
  {arXiv:0708.1595 [gr-qc]} \BibitemShut {NoStop}%
\bibitem [{\citenamefont {Ashtekar}\ \emph {et~al.}(2010)\citenamefont
  {Ashtekar}, \citenamefont {Campiglia},\ and\ \citenamefont
  {Henderson}}]{Ashtekar:2010ve}%
  \BibitemOpen
  \bibfield  {author} {\bibinfo {author} {\bibfnamefont {Abhay}\ \bibnamefont
  {Ashtekar}}, \bibinfo {author} {\bibfnamefont {Miguel}\ \bibnamefont
  {Campiglia}}, \ and\ \bibinfo {author} {\bibfnamefont {Adam}\ \bibnamefont
  {Henderson}},\ }\bibfield  {title} {\enquote {\bibinfo {title} {{Casting Loop
  Quantum Cosmology in the Spin Foam Paradigm}},}\ }\href {\doibase
  10.1088/0264-9381/27/13/135020} {\bibfield  {journal} {\bibinfo  {journal}
  {Class. Quant. Grav.}\ }\textbf {\bibinfo {volume} {27}},\ \bibinfo {pages}
  {135020} (\bibinfo {year} {2010})},\ \Eprint {http://arxiv.org/abs/1001.5147}
  {arXiv:1001.5147 [gr-qc]} \BibitemShut {NoStop}%
\bibitem [{\citenamefont {Misner}(1957)}]{Misner:1957wq}%
  \BibitemOpen
  \bibfield  {author} {\bibinfo {author} {\bibfnamefont {Charles~W.}\
  \bibnamefont {Misner}},\ }\bibfield  {title} {\enquote {\bibinfo {title}
  {{Feynman quantization of general relativity}},}\ }\href {\doibase
  10.1103/RevModPhys.29.497} {\bibfield  {journal} {\bibinfo  {journal} {Rev.
  Mod. Phys.}\ }\textbf {\bibinfo {volume} {29}},\ \bibinfo {pages} {497--509}
  (\bibinfo {year} {1957})}\BibitemShut {NoStop}%
\bibitem [{\citenamefont {Hawking}(1978)}]{Hawking:1979zw}%
  \BibitemOpen
  \bibfield  {author} {\bibinfo {author} {\bibfnamefont {S.~W.}\ \bibnamefont
  {Hawking}},\ }\bibfield  {title} {\enquote {\bibinfo {title} {{Space-Time
  Foam}},}\ }\href {\doibase 10.1016/0550-3213(78)90375-9} {\bibfield
  {journal} {\bibinfo  {journal} {Nucl. Phys.}\ }\textbf {\bibinfo {volume}
  {B144}},\ \bibinfo {pages} {349--362} (\bibinfo {year} {1978})}\BibitemShut
  {NoStop}%
\bibitem [{\citenamefont {Ding}\ and\ \citenamefont
  {Rovelli}(2010)}]{Ding:2010ye}%
  \BibitemOpen
  \bibfield  {author} {\bibinfo {author} {\bibfnamefont {You}\ \bibnamefont
  {Ding}}\ and\ \bibinfo {author} {\bibfnamefont {Carlo}\ \bibnamefont
  {Rovelli}},\ }\bibfield  {title} {\enquote {\bibinfo {title} {{Physical
  boundary Hilbert space and volume operator in the Lorentzian new spin-foam
  theory}},}\ }\href {\doibase 10.1088/0264-9381/27/20/205003} {\bibfield
  {journal} {\bibinfo  {journal} {Class. Quant. Grav.}\ }\textbf {\bibinfo
  {volume} {27}},\ \bibinfo {pages} {205003} (\bibinfo {year} {2010})},\
  \Eprint {http://arxiv.org/abs/1006.1294} {arXiv:1006.1294 [gr-qc]}
  \BibitemShut {NoStop}%
\bibitem [{\citenamefont {Rovelli}(2006)}]{Rovelli:2005yj}%
  \BibitemOpen
  \bibfield  {author} {\bibinfo {author} {\bibfnamefont {Carlo}\ \bibnamefont
  {Rovelli}},\ }\bibfield  {title} {\enquote {\bibinfo {title} {{Graviton
  propagator from background-independent quantum gravity}},}\ }\href {\doibase
  10.1103/PhysRevLett.97.151301} {\bibfield  {journal} {\bibinfo  {journal}
  {Phys. Rev. Lett.}\ }\textbf {\bibinfo {volume} {97}},\ \bibinfo {pages}
  {151301} (\bibinfo {year} {2006})},\ \Eprint
  {http://arxiv.org/abs/gr-qc/0508124} {arXiv:gr-qc/0508124} \BibitemShut
  {NoStop}%
\bibitem [{\citenamefont {Bianchi}\ \emph {et~al.}(2006)\citenamefont
  {Bianchi}, \citenamefont {Modesto}, \citenamefont {Rovelli},\ and\
  \citenamefont {Speziale}}]{Bianchi:2006uf}%
  \BibitemOpen
  \bibfield  {author} {\bibinfo {author} {\bibfnamefont {Eugenio}\ \bibnamefont
  {Bianchi}}, \bibinfo {author} {\bibfnamefont {Leonardo}\ \bibnamefont
  {Modesto}}, \bibinfo {author} {\bibfnamefont {Carlo}\ \bibnamefont
  {Rovelli}}, \ and\ \bibinfo {author} {\bibfnamefont {Simone}\ \bibnamefont
  {Speziale}},\ }\bibfield  {title} {\enquote {\bibinfo {title} {{Graviton
  propagator in loop quantum gravity}},}\ }\href {\doibase
  10.1088/0264-9381/23/23/024} {\bibfield  {journal} {\bibinfo  {journal}
  {Class. Quant. Grav.}\ }\textbf {\bibinfo {volume} {23}},\ \bibinfo {pages}
  {6989--7028} (\bibinfo {year} {2006})},\ \Eprint
  {http://arxiv.org/abs/gr-qc/0604044} {arXiv:gr-qc/0604044} \BibitemShut
  {NoStop}%
\bibitem [{\citenamefont {Alesci}\ and\ \citenamefont
  {Rovelli}(2007)}]{Alesci:2007tx}%
  \BibitemOpen
  \bibfield  {author} {\bibinfo {author} {\bibfnamefont {Emanuele}\
  \bibnamefont {Alesci}}\ and\ \bibinfo {author} {\bibfnamefont {Carlo}\
  \bibnamefont {Rovelli}},\ }\bibfield  {title} {\enquote {\bibinfo {title}
  {{The complete LQG propagator: I. Difficulties with the Barrett-Crane
  vertex}},}\ }\href {\doibase 10.1103/PhysRevD.76.104012} {\bibfield
  {journal} {\bibinfo  {journal} {Phys. Rev.}\ }\textbf {\bibinfo {volume}
  {D76}},\ \bibinfo {pages} {104012} (\bibinfo {year} {2007})},\ \Eprint
  {http://arxiv.org/abs/0708.0883} {arXiv:0708.0883 [gr-qc]} \BibitemShut
  {NoStop}%
\bibitem [{\citenamefont {Bianchi}\ \emph {et~al.}(2009)\citenamefont
  {Bianchi}, \citenamefont {Magliaro},\ and\ \citenamefont
  {Perini}}]{Bianchi:2009ri}%
  \BibitemOpen
  \bibfield  {author} {\bibinfo {author} {\bibfnamefont {Eugenio}\ \bibnamefont
  {Bianchi}}, \bibinfo {author} {\bibfnamefont {Elena}\ \bibnamefont
  {Magliaro}}, \ and\ \bibinfo {author} {\bibfnamefont {Claudio}\ \bibnamefont
  {Perini}},\ }\bibfield  {title} {\enquote {\bibinfo {title} {{LQG propagator
  from the new spin foams}},}\ }\href {\doibase
  10.1016/j.nuclphysb.2009.07.016} {\bibfield  {journal} {\bibinfo  {journal}
  {Nucl. Phys.}\ }\textbf {\bibinfo {volume} {B822}},\ \bibinfo {pages}
  {245--269} (\bibinfo {year} {2009})},\ \Eprint
  {http://arxiv.org/abs/0905.4082} {arXiv:0905.4082 [gr-qc]} \BibitemShut
  {NoStop}%
\bibitem [{\citenamefont {Bianchi}\ \emph
  {et~al.}(2010{\natexlab{b}})\citenamefont {Bianchi}, \citenamefont {Regoli},\
  and\ \citenamefont {Rovelli}}]{Bianchi:2010fj}%
  \BibitemOpen
  \bibfield  {author} {\bibinfo {author} {\bibfnamefont {Eugenio}\ \bibnamefont
  {Bianchi}}, \bibinfo {author} {\bibfnamefont {Daniele}\ \bibnamefont
  {Regoli}}, \ and\ \bibinfo {author} {\bibfnamefont {Carlo}\ \bibnamefont
  {Rovelli}},\ }\bibfield  {title} {\enquote {\bibinfo {title} {{Face amplitude
  of spinfoam quantum gravity}},}\ }\href {\doibase
  10.1088/0264-9381/27/18/185009} {\bibfield  {journal} {\bibinfo  {journal}
  {Class. Quant. Grav.}\ }\textbf {\bibinfo {volume} {27}},\ \bibinfo {pages}
  {185009} (\bibinfo {year} {2010}{\natexlab{b}})},\ \Eprint
  {http://arxiv.org/abs/1005.0764} {arXiv:1005.0764 [gr-qc]} \BibitemShut
  {NoStop}%
\bibitem [{\citenamefont {Kaminski}\ \emph
  {et~al.}(2010{\natexlab{a}})\citenamefont {Kaminski}, \citenamefont
  {Kisielowski},\ and\ \citenamefont {Lewandowski}}]{Kaminski:2009fm}%
  \BibitemOpen
  \bibfield  {author} {\bibinfo {author} {\bibfnamefont {Wojciech}\
  \bibnamefont {Kaminski}}, \bibinfo {author} {\bibfnamefont {Marcin}\
  \bibnamefont {Kisielowski}}, \ and\ \bibinfo {author} {\bibfnamefont {Jerzy}\
  \bibnamefont {Lewandowski}},\ }\bibfield  {title} {\enquote {\bibinfo {title}
  {{Spin-Foams for All Loop Quantum Gravity}},}\ }\href {\doibase
  10.1088/0264-9381/27/9/095006} {\bibfield  {journal} {\bibinfo  {journal}
  {Class. Quant. Grav.}\ }\textbf {\bibinfo {volume} {27}},\ \bibinfo {pages}
  {095006} (\bibinfo {year} {2010}{\natexlab{a}})},\ \Eprint
  {http://arxiv.org/abs/0909.0939} {arXiv:0909.0939 [gr-qc]} \BibitemShut
  {NoStop}%
\bibitem [{\citenamefont {Kaminski}\ \emph
  {et~al.}(2010{\natexlab{b}})\citenamefont {Kaminski}, \citenamefont
  {Kisielowski},\ and\ \citenamefont {Lewandowski}}]{Kaminski:2009cc}%
  \BibitemOpen
  \bibfield  {author} {\bibinfo {author} {\bibfnamefont {Wojciech}\
  \bibnamefont {Kaminski}}, \bibinfo {author} {\bibfnamefont {Marcin}\
  \bibnamefont {Kisielowski}}, \ and\ \bibinfo {author} {\bibfnamefont {Jerzy}\
  \bibnamefont {Lewandowski}},\ }\bibfield  {title} {\enquote {\bibinfo {title}
  {{The EPRL intertwiners and corrected partition function}},}\ }\href
  {\doibase 10.1088/0264-9381/27/16/165020} {\bibfield  {journal} {\bibinfo
  {journal} {Class. Quant. Grav.}\ }\textbf {\bibinfo {volume} {27}},\ \bibinfo
  {pages} {165020} (\bibinfo {year} {2010}{\natexlab{b}})},\ \Eprint
  {http://arxiv.org/abs/0912.0540} {arXiv:0912.0540 [gr-qc]} \BibitemShut
  {NoStop}%
\bibitem [{\citenamefont {Baez}(1998)}]{Baez:1997zt}%
  \BibitemOpen
  \bibfield  {author} {\bibinfo {author} {\bibfnamefont {John~C.}\ \bibnamefont
  {Baez}},\ }\bibfield  {title} {\enquote {\bibinfo {title} {{Spin foam
  models}},}\ }\href {\doibase 10.1088/0264-9381/15/7/004} {\bibfield
  {journal} {\bibinfo  {journal} {Class. Quant. Grav.}\ }\textbf {\bibinfo
  {volume} {15}},\ \bibinfo {pages} {1827--1858} (\bibinfo {year} {1998})},\
  \Eprint {http://arxiv.org/abs/gr-qc/9709052} {arXiv:gr-qc/9709052}
  \BibitemShut {NoStop}%
\bibitem [{\citenamefont {Bojowald}\ and\ \citenamefont
  {Perez}(2010)}]{Bojowald:2009im}%
  \BibitemOpen
  \bibfield  {author} {\bibinfo {author} {\bibfnamefont {Martin}\ \bibnamefont
  {Bojowald}}\ and\ \bibinfo {author} {\bibfnamefont {Alejandro}\ \bibnamefont
  {Perez}},\ }\bibfield  {title} {\enquote {\bibinfo {title} {{Spin foam
  quantization and anomalies}},}\ }\href {\doibase 10.1007/s10714-009-0892-9}
  {\bibfield  {journal} {\bibinfo  {journal} {Gen. Rel. Grav.}\ }\textbf
  {\bibinfo {volume} {42}},\ \bibinfo {pages} {877--907} (\bibinfo {year}
  {2010})},\ \Eprint {http://arxiv.org/abs/gr-qc/0303026} {arXiv:gr-qc/0303026}
  \BibitemShut {NoStop}%
\bibitem [{\citenamefont {Ashtekar}(1986)}]{Ashtekar:1986yd}%
  \BibitemOpen
  \bibfield  {author} {\bibinfo {author} {\bibfnamefont {Abhay}\ \bibnamefont
  {Ashtekar}},\ }\bibfield  {title} {\enquote {\bibinfo {title} {{New Variables
  for Classical and Quantum Gravity}},}\ }\href {\doibase
  10.1103/PhysRevLett.57.2244} {\bibfield  {journal} {\bibinfo  {journal}
  {Phys. Rev. Lett.}\ }\textbf {\bibinfo {volume} {57}},\ \bibinfo {pages}
  {2244--2247} (\bibinfo {year} {1986})}\BibitemShut {NoStop}%
\bibitem [{\citenamefont {Rovelli}\ and\ \citenamefont
  {Smolin}(1990)}]{Rovelli:1989za}%
  \BibitemOpen
  \bibfield  {author} {\bibinfo {author} {\bibfnamefont {Carlo}\ \bibnamefont
  {Rovelli}}\ and\ \bibinfo {author} {\bibfnamefont {Lee}\ \bibnamefont
  {Smolin}},\ }\bibfield  {title} {\enquote {\bibinfo {title} {{Loop Space
  Representation of Quantum General Relativity}},}\ }\href {\doibase
  10.1016/0550-3213(90)90019-A} {\bibfield  {journal} {\bibinfo  {journal}
  {Nucl. Phys.}\ }\textbf {\bibinfo {volume} {B331}},\ \bibinfo {pages} {80}
  (\bibinfo {year} {1990})}\BibitemShut {NoStop}%
\bibitem [{\citenamefont {Ashtekar}\ and\ \citenamefont
  {Isham}(1992)}]{Ashtekar:1991kc}%
  \BibitemOpen
  \bibfield  {author} {\bibinfo {author} {\bibfnamefont {Abhay}\ \bibnamefont
  {Ashtekar}}\ and\ \bibinfo {author} {\bibfnamefont {C.~J.}\ \bibnamefont
  {Isham}},\ }\bibfield  {title} {\enquote {\bibinfo {title} {{Representations
  of the holonomy algebras of gravity and nonAbelian gauge theories}},}\ }\href
  {\doibase 10.1088/0264-9381/9/6/004} {\bibfield  {journal} {\bibinfo
  {journal} {Class. Quant. Grav.}\ }\textbf {\bibinfo {volume} {9}},\ \bibinfo
  {pages} {1433--1468} (\bibinfo {year} {1992})},\ \Eprint
  {http://arxiv.org/abs/hep-th/9202053} {arXiv:hep-th/9202053} \BibitemShut
  {NoStop}%
\bibitem [{\citenamefont {Rovelli}()}]{Rovelli:2004tv}%
  \BibitemOpen
  \bibfield  {author} {\bibinfo {author} {\bibfnamefont {Carlo}\ \bibnamefont
  {Rovelli}},\ }\bibfield  {title} {\enquote {\bibinfo {title} {{Quantum
  gravity}},}\ }\href@noop {} {\ }\bibinfo {note} {Cambridge, UK: Univ. Pr.
  (2004) 455 p}\BibitemShut {NoStop}%
\bibitem [{\citenamefont {Thiemann}(2001)}]{Thiemann:2007zz}%
  \BibitemOpen
  \bibfield  {author} {\bibinfo {author} {\bibfnamefont {Thomas}\ \bibnamefont
  {Thiemann}},\ }\bibfield  {title} {\enquote {\bibinfo {title} {{Modern
  canonical quantum general relativity}},}\ }\href@noop {} {\  (\bibinfo {year}
  {2001})},\ \Eprint {http://arxiv.org/abs/gr-qc/0110034} {arXiv:gr-qc/0110034}
  \BibitemShut {NoStop}%
\bibitem [{\citenamefont {Ashtekar}\ and\ \citenamefont
  {Lewandowski}(1993)}]{Ashtekar:1993wf}%
  \BibitemOpen
  \bibfield  {author} {\bibinfo {author} {\bibfnamefont {Abhay}\ \bibnamefont
  {Ashtekar}}\ and\ \bibinfo {author} {\bibfnamefont {Jerzy}\ \bibnamefont
  {Lewandowski}},\ }\bibfield  {title} {\enquote {\bibinfo {title}
  {{Representation theory of analytic holonomy C* algebras}},}\ }\href@noop {}
  {\  (\bibinfo {year} {1993})},\ \Eprint {http://arxiv.org/abs/gr-qc/9311010}
  {arXiv:gr-qc/9311010} \BibitemShut {NoStop}%
\bibitem [{\citenamefont {Fairbairn}\ and\ \citenamefont
  {Rovelli}(2004)}]{Fairbairn:2004qe}%
  \BibitemOpen
  \bibfield  {author} {\bibinfo {author} {\bibfnamefont {Winston}\ \bibnamefont
  {Fairbairn}}\ and\ \bibinfo {author} {\bibfnamefont {Carlo}\ \bibnamefont
  {Rovelli}},\ }\bibfield  {title} {\enquote {\bibinfo {title} {{Separable
  Hilbert space in loop quantum gravity}},}\ }\href {\doibase
  10.1063/1.1763247} {\bibfield  {journal} {\bibinfo  {journal} {J. Math.
  Phys.}\ }\textbf {\bibinfo {volume} {45}},\ \bibinfo {pages} {2802--2814}
  (\bibinfo {year} {2004})},\ \Eprint {http://arxiv.org/abs/gr-qc/0403047}
  {arXiv:gr-qc/0403047} \BibitemShut {NoStop}%
\bibitem [{\citenamefont {Barbero~G.}(1995)}]{Barbero:1994ap}%
  \BibitemOpen
  \bibfield  {author} {\bibinfo {author} {\bibfnamefont {J.~Fernando}\
  \bibnamefont {Barbero~G.}},\ }\bibfield  {title} {\enquote {\bibinfo {title}
  {{Real Ashtekar variables for Lorentzian signature space times}},}\ }\href
  {\doibase 10.1103/PhysRevD.51.5507} {\bibfield  {journal} {\bibinfo
  {journal} {Phys. Rev.}\ }\textbf {\bibinfo {volume} {D51}},\ \bibinfo {pages}
  {5507--5510} (\bibinfo {year} {1995})},\ \Eprint
  {http://arxiv.org/abs/gr-qc/9410014} {arXiv:gr-qc/9410014} \BibitemShut
  {NoStop}%
\bibitem [{\citenamefont {Immirzi}(1997)}]{Immirzi:1996di}%
  \BibitemOpen
  \bibfield  {author} {\bibinfo {author} {\bibfnamefont {Giorgio}\ \bibnamefont
  {Immirzi}},\ }\bibfield  {title} {\enquote {\bibinfo {title} {{Real and
  complex connections for canonical gravity}},}\ }\href {\doibase
  10.1088/0264-9381/14/10/002} {\bibfield  {journal} {\bibinfo  {journal}
  {Class. Quant. Grav.}\ }\textbf {\bibinfo {volume} {14}},\ \bibinfo {pages}
  {L177--L181} (\bibinfo {year} {1997})},\ \Eprint
  {http://arxiv.org/abs/gr-qc/9612030} {arXiv:gr-qc/9612030} \BibitemShut
  {NoStop}%
\bibitem [{\citenamefont {Oriti}(2009)}]{Oriti:2009wn}%
  \BibitemOpen
  \bibfield  {author} {\bibinfo {author} {\bibfnamefont {Daniele}\ \bibnamefont
  {Oriti}},\ }\bibfield  {title} {\enquote {\bibinfo {title} {{The group field
  theory approach to quantum gravity: some recent results}},}\ }\href@noop {}
  {\  (\bibinfo {year} {2009})},\ \Eprint {http://arxiv.org/abs/0912.2441}
  {arXiv:0912.2441 [hep-th]} \BibitemShut {NoStop}%
\bibitem [{\citenamefont {Perini}\ \emph {et~al.}(2009)\citenamefont {Perini},
  \citenamefont {Rovelli},\ and\ \citenamefont {Speziale}}]{Perini:2008pd}%
  \BibitemOpen
  \bibfield  {author} {\bibinfo {author} {\bibfnamefont {Claudio}\ \bibnamefont
  {Perini}}, \bibinfo {author} {\bibfnamefont {Carlo}\ \bibnamefont {Rovelli}},
  \ and\ \bibinfo {author} {\bibfnamefont {Simone}\ \bibnamefont {Speziale}},\
  }\bibfield  {title} {\enquote {\bibinfo {title} {{Self-energy and vertex
  radiative corrections in LQG}},}\ }\href {\doibase
  10.1016/j.physletb.2009.10.076} {\bibfield  {journal} {\bibinfo  {journal}
  {Phys. Lett.}\ }\textbf {\bibinfo {volume} {B682}},\ \bibinfo {pages}
  {78--84} (\bibinfo {year} {2009})},\ \Eprint {http://arxiv.org/abs/0810.1714}
  {arXiv:0810.1714 [gr-qc]} \BibitemShut {NoStop}%
\bibitem [{\citenamefont {Freidel}\ \emph {et~al.}(2009)\citenamefont
  {Freidel}, \citenamefont {Gurau},\ and\ \citenamefont
  {Oriti}}]{Freidel:2009hd}%
  \BibitemOpen
  \bibfield  {author} {\bibinfo {author} {\bibfnamefont {Laurent}\ \bibnamefont
  {Freidel}}, \bibinfo {author} {\bibfnamefont {Razvan}\ \bibnamefont {Gurau}},
  \ and\ \bibinfo {author} {\bibfnamefont {Daniele}\ \bibnamefont {Oriti}},\
  }\bibfield  {title} {\enquote {\bibinfo {title} {{Group field theory
  renormalization - the 3d case: power counting of divergences}},}\ }\href
  {\doibase 10.1103/PhysRevD.80.044007} {\bibfield  {journal} {\bibinfo
  {journal} {Phys. Rev.}\ }\textbf {\bibinfo {volume} {D80}},\ \bibinfo {pages}
  {044007} (\bibinfo {year} {2009})},\ \Eprint {http://arxiv.org/abs/0905.3772}
  {arXiv:0905.3772 [hep-th]} \BibitemShut {NoStop}%
\bibitem [{\citenamefont {Bonzom}\ and\ \citenamefont
  {Smerlak}(2010)}]{Bonzom:2010zh}%
  \BibitemOpen
  \bibfield  {author} {\bibinfo {author} {\bibfnamefont {Valentin}\
  \bibnamefont {Bonzom}}\ and\ \bibinfo {author} {\bibfnamefont {Matteo}\
  \bibnamefont {Smerlak}},\ }\bibfield  {title} {\enquote {\bibinfo {title}
  {{Bubble divergences from twisted cohomology}},}\ }\href@noop {} {\
  (\bibinfo {year} {2010})},\ \Eprint {http://arxiv.org/abs/1008.1476}
  {arXiv:1008.1476 [math-ph]} \BibitemShut {NoStop}%
\bibitem [{\citenamefont {Krajewski}\ \emph {et~al.}(2010)\citenamefont
  {Krajewski}, \citenamefont {Magnen}, \citenamefont {Rivasseau}, \citenamefont
  {Tanasa},\ and\ \citenamefont {Vitale}}]{Krajewski:2010yq}%
  \BibitemOpen
  \bibfield  {author} {\bibinfo {author} {\bibfnamefont {Thomas}\ \bibnamefont
  {Krajewski}}, \bibinfo {author} {\bibfnamefont {Jacques}\ \bibnamefont
  {Magnen}}, \bibinfo {author} {\bibfnamefont {Vincent}\ \bibnamefont
  {Rivasseau}}, \bibinfo {author} {\bibfnamefont {Adrian}\ \bibnamefont
  {Tanasa}}, \ and\ \bibinfo {author} {\bibfnamefont {Patrizia}\ \bibnamefont
  {Vitale}},\ }\bibfield  {title} {\enquote {\bibinfo {title} {{Quantum
  Corrections in the Group Field Theory Formulation of the EPRL/FK Models}},}\
  }\href@noop {} {\  (\bibinfo {year} {2010})},\ \Eprint
  {http://arxiv.org/abs/1007.3150} {arXiv:1007.3150 [gr-qc]} \BibitemShut
  {NoStop}%
\bibitem [{\citenamefont {Oeckl}(2003{\natexlab{a}})}]{Oeckl:2003vu}%
  \BibitemOpen
  \bibfield  {author} {\bibinfo {author} {\bibfnamefont {Robert}\ \bibnamefont
  {Oeckl}},\ }\bibfield  {title} {\enquote {\bibinfo {title} {{A 'general
  boundary' formulation for quantum mechanics and quantum gravity}},}\ }\href
  {\doibase 10.1016/j.physletb.2003.08.043} {\bibfield  {journal} {\bibinfo
  {journal} {Phys. Lett.}\ }\textbf {\bibinfo {volume} {B575}},\ \bibinfo
  {pages} {318--324} (\bibinfo {year} {2003}{\natexlab{a}})},\ \Eprint
  {http://arxiv.org/abs/hep-th/0306025} {arXiv:hep-th/0306025} \BibitemShut
  {NoStop}%
\bibitem [{\citenamefont {Regge}(1961)}]{Regge:1961px}%
  \BibitemOpen
  \bibfield  {author} {\bibinfo {author} {\bibfnamefont {T.}~\bibnamefont
  {Regge}},\ }\bibfield  {title} {\enquote {\bibinfo {title} {{General
  relativity without coordinates}},}\ }\href {\doibase 10.1007/BF02733251}
  {\bibfield  {journal} {\bibinfo  {journal} {Nuovo Cim.}\ }\textbf {\bibinfo
  {volume} {19}},\ \bibinfo {pages} {558--571} (\bibinfo {year}
  {1961})}\BibitemShut {NoStop}%
\bibitem [{\citenamefont {Ponzano}\ and\ \citenamefont
  {Regge}()}]{PonzanoRegge:1968}%
  \BibitemOpen
  \bibfield  {author} {\bibinfo {author} {\bibfnamefont {G.}~\bibnamefont
  {Ponzano}}\ and\ \bibinfo {author} {\bibfnamefont {T.}~\bibnamefont
  {Regge}},\ }\bibfield  {title} {\enquote {\bibinfo {title} {Semiclassical
  limit of racah coeffecients},}\ }\href@noop {} {\ }\bibinfo {note}
  {Spectroscopic and Group Theoretical Methods in Physics, edited by F.Block
  (North Holland, Amsterdam, 1968)}\BibitemShut {NoStop}%
\bibitem [{\citenamefont {Boulatov}(1992)}]{Boulatov:1992vp}%
  \BibitemOpen
  \bibfield  {author} {\bibinfo {author} {\bibfnamefont {D.~V.}\ \bibnamefont
  {Boulatov}},\ }\bibfield  {title} {\enquote {\bibinfo {title} {{A Model of
  three-dimensional lattice gravity}},}\ }\href {\doibase
  10.1142/S0217732392001324} {\bibfield  {journal} {\bibinfo  {journal} {Mod.
  Phys. Lett.}\ }\textbf {\bibinfo {volume} {A7}},\ \bibinfo {pages}
  {1629--1646} (\bibinfo {year} {1992})},\ \Eprint
  {http://arxiv.org/abs/hep-th/9202074} {arXiv:hep-th/9202074} \BibitemShut
  {NoStop}%
\bibitem [{\citenamefont {Noui}\ and\ \citenamefont
  {Perez}(2005)}]{Noui:2004iy}%
  \BibitemOpen
  \bibfield  {author} {\bibinfo {author} {\bibfnamefont {Karim}\ \bibnamefont
  {Noui}}\ and\ \bibinfo {author} {\bibfnamefont {Alejandro}\ \bibnamefont
  {Perez}},\ }\bibfield  {title} {\enquote {\bibinfo {title} {{Three
  dimensional loop quantum gravity: Physical scalar product and spin foam
  models}},}\ }\href {\doibase 10.1088/0264-9381/22/9/017} {\bibfield
  {journal} {\bibinfo  {journal} {Class. Quant. Grav.}\ }\textbf {\bibinfo
  {volume} {22}},\ \bibinfo {pages} {1739--1762} (\bibinfo {year} {2005})},\
  \Eprint {http://arxiv.org/abs/gr-qc/0402110} {arXiv:gr-qc/0402110}
  \BibitemShut {NoStop}%
\bibitem [{\citenamefont {Geloun}\ \emph {et~al.}(2010)\citenamefont {Geloun},
  \citenamefont {Gurau},\ and\ \citenamefont {Rivasseau}}]{Geloun:2010vj}%
  \BibitemOpen
  \bibfield  {author} {\bibinfo {author} {\bibfnamefont {Joseph~Ben}\
  \bibnamefont {Geloun}}, \bibinfo {author} {\bibfnamefont {Razvan}\
  \bibnamefont {Gurau}}, \ and\ \bibinfo {author} {\bibfnamefont {Vincent}\
  \bibnamefont {Rivasseau}},\ }\bibfield  {title} {\enquote {\bibinfo {title}
  {{EPRL/FK Group Field Theory}},}\ }\href@noop {} {\  (\bibinfo {year}
  {2010})},\ \Eprint {http://arxiv.org/abs/1008.0354} {arXiv:1008.0354
  [hep-th]} \BibitemShut {NoStop}%
\bibitem [{\citenamefont {Dupuis}\ and\ \citenamefont
  {Livine}(2010)}]{Dupuis:2010jn}%
  \BibitemOpen
  \bibfield  {author} {\bibinfo {author} {\bibfnamefont {Maite}\ \bibnamefont
  {Dupuis}}\ and\ \bibinfo {author} {\bibfnamefont {Etera~R.}\ \bibnamefont
  {Livine}},\ }\bibfield  {title} {\enquote {\bibinfo {title} {{Lifting SU(2)
  Spin Networks to Projected Spin Networks}},}\ }\href {\doibase
  10.1103/PhysRevD.82.064044} {\bibfield  {journal} {\bibinfo  {journal} {Phys.
  Rev.}\ }\textbf {\bibinfo {volume} {D82}},\ \bibinfo {pages} {064044}
  (\bibinfo {year} {2010})},\ \Eprint {http://arxiv.org/abs/1008.4093}
  {arXiv:1008.4093 [gr-qc]} \BibitemShut {NoStop}%
\bibitem [{\citenamefont {Ooguri}(1992)}]{Ooguri:1992eb}%
  \BibitemOpen
  \bibfield  {author} {\bibinfo {author} {\bibfnamefont {Hirosi}\ \bibnamefont
  {Ooguri}},\ }\bibfield  {title} {\enquote {\bibinfo {title} {{Topological
  lattice models in four-dimensions}},}\ }\href {\doibase
  10.1142/S0217732392004171} {\bibfield  {journal} {\bibinfo  {journal} {Mod.
  Phys. Lett.}\ }\textbf {\bibinfo {volume} {A7}},\ \bibinfo {pages}
  {2799--2810} (\bibinfo {year} {1992})},\ \Eprint
  {http://arxiv.org/abs/hep-th/9205090} {arXiv:hep-th/9205090} \BibitemShut
  {NoStop}%
\bibitem [{\citenamefont {Barbieri}(1998)}]{Barbieri:1997ks}%
  \BibitemOpen
  \bibfield  {author} {\bibinfo {author} {\bibfnamefont {A.}~\bibnamefont
  {Barbieri}},\ }\bibfield  {title} {\enquote {\bibinfo {title} {{Quantum
  tetrahedra and simplicial spin networks}},}\ }\href {\doibase
  10.1016/S0550-3213(98)00093-5} {\bibfield  {journal} {\bibinfo  {journal}
  {Nucl. Phys.}\ }\textbf {\bibinfo {volume} {B518}},\ \bibinfo {pages}
  {714--728} (\bibinfo {year} {1998})},\ \Eprint
  {http://arxiv.org/abs/gr-qc/9707010} {arXiv:gr-qc/9707010} \BibitemShut
  {NoStop}%
\bibitem [{\citenamefont {Barrett}\ and\ \citenamefont
  {Crane}(1998)}]{Barrett:1997gw}%
  \BibitemOpen
  \bibfield  {author} {\bibinfo {author} {\bibfnamefont {John~W.}\ \bibnamefont
  {Barrett}}\ and\ \bibinfo {author} {\bibfnamefont {Louis}\ \bibnamefont
  {Crane}},\ }\bibfield  {title} {\enquote {\bibinfo {title} {{Relativistic
  spin networks and quantum gravity}},}\ }\href {\doibase 10.1063/1.532254}
  {\bibfield  {journal} {\bibinfo  {journal} {J. Math. Phys.}\ }\textbf
  {\bibinfo {volume} {39}},\ \bibinfo {pages} {3296--3302} (\bibinfo {year}
  {1998})},\ \Eprint {http://arxiv.org/abs/gr-qc/9709028} {arXiv:gr-qc/9709028}
  \BibitemShut {NoStop}%
\bibitem [{\citenamefont {Oriti}(2006)}]{Oriti:2006se}%
  \BibitemOpen
  \bibfield  {author} {\bibinfo {author} {\bibfnamefont {Daniele}\ \bibnamefont
  {Oriti}},\ }\bibfield  {title} {\enquote {\bibinfo {title} {{The group field
  theory approach to quantum gravity}},}\ }\href@noop {} {\  (\bibinfo {year}
  {2006})},\ \Eprint {http://arxiv.org/abs/gr-qc/0607032} {arXiv:gr-qc/0607032}
  \BibitemShut {NoStop}%
\bibitem [{\citenamefont {Ambjorn}\ \emph {et~al.}(1993)\citenamefont
  {Ambjorn}, \citenamefont {Jurkiewicz},\ and\ \citenamefont
  {Kristjansen}}]{Ambjorn:1992aw}%
  \BibitemOpen
  \bibfield  {author} {\bibinfo {author} {\bibfnamefont {Jan}\ \bibnamefont
  {Ambjorn}}, \bibinfo {author} {\bibfnamefont {Jerzy}\ \bibnamefont
  {Jurkiewicz}}, \ and\ \bibinfo {author} {\bibfnamefont {Charlotte~F.}\
  \bibnamefont {Kristjansen}},\ }\bibfield  {title} {\enquote {\bibinfo {title}
  {{Quantum gravity, dynamical triangulations and higher derivative
  regularization}},}\ }\href {\doibase 10.1016/0550-3213(93)90075-Z} {\bibfield
   {journal} {\bibinfo  {journal} {Nucl. Phys.}\ }\textbf {\bibinfo {volume}
  {B393}},\ \bibinfo {pages} {601--632} (\bibinfo {year} {1993})},\ \Eprint
  {http://arxiv.org/abs/hep-th/9208032} {arXiv:hep-th/9208032} \BibitemShut
  {NoStop}%
\bibitem [{\citenamefont {Oeckl}(2003{\natexlab{b}})}]{Oeckl:2002ia}%
  \BibitemOpen
  \bibfield  {author} {\bibinfo {author} {\bibfnamefont {Robert}\ \bibnamefont
  {Oeckl}},\ }\bibfield  {title} {\enquote {\bibinfo {title} {{Renormalization
  of discrete models without background}},}\ }\href {\doibase
  10.1016/S0550-3213(03)00145-7} {\bibfield  {journal} {\bibinfo  {journal}
  {Nucl. Phys.}\ }\textbf {\bibinfo {volume} {B657}},\ \bibinfo {pages}
  {107--138} (\bibinfo {year} {2003}{\natexlab{b}})},\ \Eprint
  {http://arxiv.org/abs/gr-qc/0212047} {arXiv:gr-qc/0212047} \BibitemShut
  {NoStop}%
\bibitem [{\citenamefont {Rovelli}(2010)}]{Rovelli:2010vv}%
  \BibitemOpen
  \bibfield  {author} {\bibinfo {author} {\bibfnamefont {Carlo}\ \bibnamefont
  {Rovelli}},\ }\bibfield  {title} {\enquote {\bibinfo {title} {{Simple model
  for quantum general relativity from loop quantum gravity}},}\ }\href@noop {}
  {\  (\bibinfo {year} {2010})},\ \Eprint {http://arxiv.org/abs/1010.1939}
  {arXiv:1010.1939 [gr-qc]} \BibitemShut {NoStop}%
\bibitem [{\citenamefont {Rovelli}\ and\ \citenamefont {Smerlak}()}]{RovSmer}%
  \BibitemOpen
  \bibfield  {author} {\bibinfo {author} {\bibfnamefont {Carlo}\ \bibnamefont
  {Rovelli}}\ and\ \bibinfo {author} {\bibfnamefont {Matteo}\ \bibnamefont
  {Smerlak}},\ }\bibfield  {title} {\enquote {\bibinfo {title} {{Summing over
  triangulations or refining the triangulation?}}}\ }\href@noop {} {\ }\bibinfo
  {note} {To appear}\BibitemShut {NoStop}%
\bibitem [{\citenamefont {Zapata}(2002)}]{Zapata:2002eu}%
  \BibitemOpen
  \bibfield  {author} {\bibinfo {author} {\bibfnamefont {Jose~A.}\ \bibnamefont
  {Zapata}},\ }\bibfield  {title} {\enquote {\bibinfo {title} {{Continuum spin
  foam model for 3d gravity}},}\ }\href {\doibase 10.1063/1.1509850} {\bibfield
   {journal} {\bibinfo  {journal} {J. Math. Phys.}\ }\textbf {\bibinfo {volume}
  {43}},\ \bibinfo {pages} {5612--5623} (\bibinfo {year} {2002})},\ \Eprint
  {http://arxiv.org/abs/gr-qc/0205037} {arXiv:gr-qc/0205037} \BibitemShut
  {NoStop}%
\end{thebibliography}%
\end{document}